\documentclass[pra,twocolumn,showpacs,superscriptaddress]{revtex4-1}
\usepackage{epsfig,graphicx,amssymb,color,bm}
\usepackage{amsmath,float}
\usepackage[colorlinks, linkcolor=blue, urlcolor=blue, citecolor=blue, unicode=true]{hyperref}
\usepackage{setspace}
\usepackage{url}
\newcommand{\eref}[1]{Eq.~(\ref{#1})}
\newcommand{\esref}[1]{Eqs.~(\ref{#1})}
\newcommand{\fref}[1]{Fig.~\ref{#1}}
\begin{document}

\title{Bistable cooling in optomechanical system}
\author{Wen-Zhao Zhang}\thanks{zhangwz@csrc.ac.cn}
\affiliation{Beijing Computational Science Research Center (CSRC), Beijing 100193, China}
\author{Wen-Lin Li}
\affiliation{School of Physics, Dalian University of Technology, Dalian 116024, China}
\author{Jiong Cheng}
\affiliation{Department of Physics, Ningbo University, Ningbo 315211, China}
\author{Qingxia Mu}
\affiliation{Mathematics and Physics Department, North China Electric Power University, Beijing 102206, China}
\begin{abstract}
A scheme is presented to optimize the optomechanical cooling of mechanical resonator in instability regime.
Based on the stability analysis, we uncovered a distinct bistable effect of photons and phonons, which can be used to realize a strong nonlinear effect even in the single-photon weak coupling regime.
Considering the experimental realization, we investigate the sideband cooling in bistable regime with and without quantum nonlinearity.
It is shown that the fluctuation of the steady state phonons can be excellently suppressed at a rather low level due to the anti-rotating-wave effect, and it does not require high quality factor of the cavity.
Our scheme offers a new perspective for optimizing the sideband cooling of mechanical resonators in the weak coupling regime.
\end{abstract}

\maketitle

\section{introduction}

Optomechanical system provides us with a convenient way to observe the quantum behavior of mesoscopic and macroscopic objects, thereby opens up a broad spectrum of applications ranging from ultrasensitive detection \cite{PhysRevLett.115.211104,PhysRevX.7.021008}, quantum information processing \cite{OE.25.010779,PhysRevA.94.033821,lpor.201600284,PhysRevApplied.4.024003} to the test of foundations of physics which includes the exploration of the elusive boundary between the quantum and the classical world \cite{PhysRevA.94.012334}, the research of the nonclassical state of mechanical oscillators \cite{PhysRevLett.101.200503}, and so on.
For most applications, it is a prerequisite to cool the mechanical resonator close to it's quantum ground state so as to suppress the thermal noise.
Recently ground state cooling of mechanical resonators has been demonstrated through various approaches including standard cryogenic cooling \cite{nature08967,nphys1303}, optical feedback cooling \cite{PhysRevLett.99.017201,PhysRevLett.80.688} and radiation-pressure cooling \cite{nature10261,PhysRevLett.110.153606}, along with many theoretical and experimental efforts on novel cooling schemes, such as cooling in non-Markovian regime \cite{PhysRevA.93.063853,PhysRevLett.99.093902},
cooling by heating \cite{PhysRevLett.108.120602,PhysRevLett.115.223602,PhysRevA.94.053836}, cooling with modulations \cite{PhysRevA.91.023818,PhysRevLett.110.153606}, and cooling by using cavity electromagnetically induced transparency \cite{PhysRevA.84.051801}.

The interaction between light and mechanical motion due to radiation pressure is intrinsically nonlinear.
A variety of interesting quantum nonlinear phenomena have been uncovered in optomechanical system, such as optomechanical Kerr nonlinearity \cite{PhysRevA.91.063836,OE.25.010779,PhysRevA.88.043826}, optomechanically induced transparency \cite{PhysRevA.95.033803} and optomechanical bistability \cite{EPL.115.64002,PhysRevA.88.043826}.
These nonlinear effects are potential resources for optimizing mechanical oscillator cooling.
Currently, experimental techniques of cavity optomechanics are still in the single-photon weak coupling regime ($g\ll\omega_m$) \cite{RevModPhys.86.1391}.
However it draws relatively few works on cooling mechanical oscillators with nonlinear effect because a prerequisite of strong nonlinearity is required \cite{PhysRevA.85.051803,PhysRevA.92.023811}.
In order to utilize the nonlinearity of the optmechanical system in mechanical oscillator cooling, one can introduce an additional nonlinear interaction potential to enhance the linearized cooling effect \cite{PhysRevLett.117.173602}.
We noticed that for the bistability of optomechanical system, strong single-photon coupling rate is not a necessary condition to investigate the nonlinear effect, such as equivalent Kerr nonlinearity \cite{PhysRevA.88.043826}, optomechanical entanglement in the bistable regime \cite{PhysRevA.84.033846} and perfect photon absorption \cite{EPL.115.64002}.
A driven optomechanical cavity may have two coexisting stable steady states, and one of the stable states exhibits strong nonlinearity even the single-photon coupling rate is rather weak.
There is a meaningful example for atomic and molecular cooling optimization in ``bad" cavities, which is related to the phenomenon of optical bi- or multi-stability \cite{PhysRevLett.99.103002}.
Thus, this provides us with the possibility to produce strong nonlinear effects in weak nonlinear conditions, which allows us to optimize the cooling process by using the nonlinear effects of the optomechanical system itself, i.e. $g a^{\dag}a(b^{\dag}+b)$.
In addition, the ground state cooling in instability (bistability) regime is also a prerequisite for studying the instable quantum behavior.

In this paper, we propose a scheme to optimize mechanical oscillator cooling in fundamental optomechanical system.
This scheme does not require the stringent condition that the single-photon optomechanical coupling strength $g$ is on the order of the mechanical resonance frequency $\omega_m$ \cite{PhysRevA.85.051803,PhysRevA.92.023811} and does not introduces auxiliary nonlinear interactions \cite{PhysRevLett.117.173602}.
Our results show that an optimized cooling can be achieved in single-photon weak coupling and low cavity quality factor regime due to the anti-rotating-wave effect caused by the bistability.
By examining the Routh-Hurwitz (RH) criterion, we exhibit the unstable characteristics of the system.
By studying the nonlinear properties of the classical part, we show that the bistable behavior of the mechanical system will cause a multivalued field-enhanced coupling rate $G$ and then affect the steady state phonon number.
Similar conclusions to exhibit bistable behavior are also studied in Ref. \cite{PhysRevA.84.033846,EPL.115.64002,PhysRevA.88.043826}.
After numerical investigation of the system, we show that quantum nonlinearity can be neglected in the discussion of bistable cooling.
In realistic experimental situation, our scheme for optimized cooling is essentially based on the classical nonlinear effects that realize the parameter modulation of phonon fluctuations in weak coupling regime.
The bistability of optomechanical system can induce strong nonlinearity in single-photon weak coupling regime, so that the field-enhanced coupling rate $G$ can be compared with the mechanical frequency $\omega_m$.
Under this conditions, rotating-wave approximation is no longer suitable while in some special parameters, the effect of the anti-rotating-wave terms can greatly optimize the cooling effect, which is shown in \fref{figadd}.
The cooling phonon number can reach $10^{-3}$ of that without considering the nonlinear effect.

\section{Model and Hamiltonian}
\begin{figure}[bt]
  \centering
  \includegraphics[width=8.5cm]{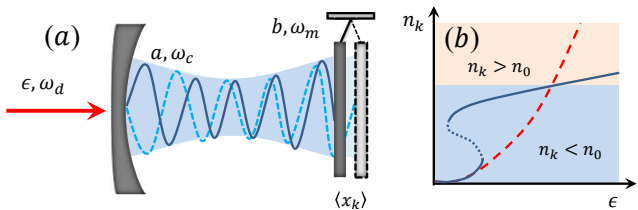}\\
  \caption{(a) Schematic diagram of a standard optomechanical system.
  The average displacement of the mechanical oscillator $\langle x_k\rangle$ will exhibits bistability behavior in instability condition.
  (b) Evolution schematic diagram of the average photon number $n_k$ as a function of driving strength $\epsilon$ in the bistable condition.
  The blue solid line represents the case of contain nonlinearity.
  The red dashed line represents the case of using linearized approximation. }\label{fig1}
\end{figure}

Here we consider a standard model of optomechanical system in which the cavity mode driven by a laser coupled to the mechanical resonator via radiation pressure, as shown in \fref{fig1}(a), the Hamiltonian reads ($\hbar=1$)
\begin{eqnarray}
  H &=& H_{sys}+H_d, \label{hmi}\\
  H_{sys} &=& \omega_c a^{\dag}a+\omega_m b^{\dag}b-g a^{\dag}a(b^{\dag}+b),\\
  H_d &=& \epsilon (a^{\dag}e^{-i \omega_d t}+ae^{i \omega_d t}),
\end{eqnarray}
where $a$ and $b$ are the bosonic operators for the optical and mechanical modes with frequencies $\omega_c$ and $\omega_m$ respectively.
The single-photon coupling coefficient of the optomechanical interaction is $g=(\omega _{c}/L)\sqrt{\hbar /(2m\omega _{m})}$.
$H_d$ describes the standard continuous-wave drive, $\omega_d$ is the angular frequency of the laser and $\epsilon$ is the cavity driving strength, given as $\epsilon \equiv 2\sqrt{P \kappa_{ex}/(\hbar \omega_d)}$, with $P$ the input power of the laser and $\kappa_{ex}$ the input rate of the cavity.
In the frame rotating at input laser frequency $\omega_d$, we can obtain the quantum Langevin equations,
\begin{eqnarray}
  \dot{a} &=& -(i\Delta+\kappa/2)a+iga(b+b^{\dag})+\sqrt{\kappa}a_{in}-i\epsilon, \\
  \dot{b} &=&-(i\omega_m+\gamma/2)b+iga^{\dag}a+\sqrt{\gamma}b_{in}.
\end{eqnarray}

The optical and mechanical field operators can be split into the classical and quantum components: $a\rightarrow\alpha+a_1$ and $b\rightarrow\beta+b_1$, where $\alpha\equiv\langle a\rangle$, $\beta\equiv \langle b \rangle$. The time evolution of the annihilation operators of the system in the Heisenberg picture are then governed by
\begin{eqnarray}
  \dot{\alpha} &=& -(i\Delta+\kappa/2)\alpha+ig\alpha(\beta+\beta^*)-i\epsilon,\nonumber\\
  \dot{\beta} &=& -(i\omega_m+\gamma/2)\beta+ig|\alpha|^2, \label{cpart}
\end{eqnarray}
\begin{eqnarray}
  \dot{a_1} &=& -(i\Delta'+\kappa/2)a_1+ig\alpha(b_1+b_1^{\dag}),\nonumber\\
  &&+iga_1(b_1+b_1^{\dag})+\sqrt{\kappa}a_{in}, \nonumber\\
  \dot{b_1} &=&-(i\omega_m+\gamma/2)b_1+ig(\alpha^*a_1+\alpha a_1^{\dag}),\nonumber\\
  &&+iga_1^{\dag}a_1+\sqrt{\gamma}b_{in},\label{qpart}
\end{eqnarray}
where $\Delta=\omega_c-\omega_d$ is the detuning between the driving field and the cavity frequency, and $\Delta'=\Delta-g(\beta+\beta^*)$ is the effective detuning of the system.
The nonlinear equations \esref{cpart} describe the dynamical evolution of the classical part of system, which is governed by the adjustable parameters, $\Delta$, $\kappa$ and $\epsilon$.
Due to the weak coupling of the single photon interaction in the experimental realization ($g\ll\{\kappa,\omega_m \}$), we can effectively ignore the nonlinear term $ig\alpha(\beta+\beta^*)$ to simplify the calculation \cite{RevModPhys.86.1391}.
Under steady-state conditions, this approximation is reasonable and usually used in the discussion of sideband cooling \cite{PhysRevA.77.033804}, optomechanical entanglement \cite{PhysRevA.84.051801}, optomechanical induced transparency \cite{PhysRevA.90.043825}, and so on.
Thus, we can get the unique steady state solution of the system according to \esref{cpart} in the stable regime, where
$\alpha_0\approx-i\epsilon/(i\Delta+\kappa/2)$, the corresponding mean photon number $n_0\approx\epsilon^2/(\Delta^2+\kappa^2/4)$, the steady state displacement $\langle x_0\rangle\approx-2gn_0/\omega_m$ \cite{PhysRevLett.110.153606}.
In addition, we also need to judge the choice of coupling strength $g$ according to the specific system to satisfy the approximation.
But, in the study of the nonlinear effects in bistability \cite{PhysRevA.91.023813,PhysRevA.93.023802}, chaos \cite{PhysRevLett.114.253601,PhysRevE.87.052906} and rebirth effect of entanglement \cite{PhysRevLett.112.110406}, the nonlinear term $ig\alpha(\beta+\beta^*)$ can be not simply ignored even that the single-photon coupling rate is rather weak.
As shown in \fref{fig1}(b) blue solid line, there is two steady state solutions for the same set of parameters (one is metastability) in the bistable regime.
As a typical bistable effect in optomechanical system \cite{PhysRevA.84.033846}, same driven strength $\epsilon$ can cause different mean steady photon number $n_k$, and the corresponding average displacement $x_k$ are also different.
There are two coexisting different dynamic behaviors for the same parameters.
Which means that there are two different cooling results in the study of the optomechanical sideband cooling.

For the quantum part, i.e. \esref{qpart}, we can introduce the so-called ``linearized" approximate description of optomechanics when the nonlinear terms $iga_1(b_1+b_1^{\dag})$ and $iga_1^{\dag}a_1$ can be omitted as being smaller by a factor $\alpha$ \cite{RevModPhys.86.1391}.
This is an efficient approximation to the study of the linearized quantum effects in optomechanical system.
Under this condition, the only correlation between the classical part and the quantum part is the field-enhanced coupling rate $G=\alpha g$, and dynamics of the quantum part have no distinguishing ability to $\alpha$ and $g$ with the same value of $G$.
Thus, the single-photon coupling rate $g$ will not affect the dynamics of the quantum part as an effective parameter.
In this case when we study the quantum effect in optomechanical system, the contribution of the classical part is only considered as an adjustable parameter.
It can be found that this kind of process simplifies calculations but ignores many quantum effects of system.
Therefor, when we consider the nonlinear effects of the optomechanical system, such as the Kerr effect \cite{PhysRevA.91.063836} and duffing effect \cite{PhysRevA.91.013834}, we need to reconsider the contribution of nonlinear terms $iga_1(b_1+b_1^{\dag})$ and $iga_1^{\dag}a_1$.

To understand the characteristic of the system in nonlinear regime better, we use Routh-Hurwitz (RH) criterion \cite{PhysRevA.35.5288} to evaluate the instability of the optomechanical system.
According to \eref{cpart}, the single-photon coupling rate $g$ is the key parameter to induce nonlinearity, and the weight in the formulas directly affects the nonlinear behavior of the system i.e. $g^2/(\omega_m \kappa)$.
On the contrary, the larger the value of the detuning frequency frequency $\Delta$ is, the smaller the effective nonlinear coupling is.
On one hand, the dissipation rate $\kappa$ will contribute an exponential attenuation to the number of photons in the cavity, which makes the system stable.
On the other hand, the driving $\epsilon$ can increase the photon number in the cavity, which makes the system unstable.
\begin{figure}[tb]
  \centering
  \includegraphics[width=8.5cm]{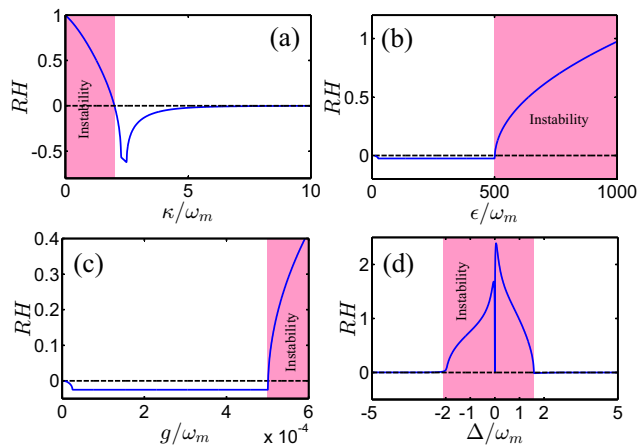}\\
  \caption{(a) RH criterion as function of $\kappa$, the coupling strength $g/\omega_m=10^{-3}$, the driving strength $\epsilon/\omega_m=10^3$, the detuning frequency $\Delta/\omega_m=1$.
  (b) RH criterion as function of $\epsilon$, the dissipation rate $\kappa/\omega_m=10^{-1}$, the coupling strength $g/\omega_m=10^{-3}$, the detuning frequency $\Delta/\omega_m=1$.
  (c) RH criterion as function of $g$, the dissipation rate $\kappa/\omega_m=10^{-1}$, the driving strength $\epsilon/\omega_m=10^3$, the detuning frequency $\Delta/\omega_m=1$.
  (d) RH criterion as function of $\Delta$, the dissipation rate $\kappa/\omega_m=10^{-1}$, the driving strength $\epsilon/\omega_m=10^3$, the coupling strength $g/\omega_m=10^{-3}$.
  Other parameters are $\gamma/\omega_m=10^{-5}$.}\label{fig2}
\end{figure}
Thus, we plot the RH criterion as a function of critical parameters $\kappa$, $\epsilon$, $g$ and $\Delta$ in \fref{fig2}.
As shown in \fref{fig2}a, the system will exhibit unstable behavior when the dissipation rate is lower than a specific value ($\kappa<2\omega_m$ for $\epsilon/\omega_m=10^3$ and $g/\omega_m=10^{-3}$).
With the increase of dissipation rate, if it exceeds a specific value $2\omega_m$, the system will become stable due to the dynamic balance of photons between ``dissipation" and ``driving".
Under this condition, dissipation will rapidly stabilize the system and make it equilibrium with the cavity environment.
As shown in \fref{fig2}b, the system will exhibit stable behavior when the driving strength is weak.
As the drive strength increases, the system exhibits an unstable behavior when it exceeds a special value $\epsilon>500\omega_m$.
At this time, laser driving will continually increase the number of photons in the cavity and eventually lead to instability.
As shown in \fref{fig2}c, the system could be regarded as a linear system when the nonlinear coupling rate $g$ is small enough.
As the coupling rate $g$ increases and exceeds a special value $g>5\times 10^{-4}\omega_m$, which is still in weak coupling regime, the system will show strong nonlinearity and its dynamics become unstable.
In \fref{fig2}d, given parameters $\kappa/\omega_m=0.1$, $\epsilon/\omega_m=10^3$, $g/\omega_m=10^{-3}$, the system exhibits unstable behavior, both the parametric down-conversion interaction $a_1b_1+a_1^{\dag}b_1^{\dag}$ on resonance (red-detuning) and the beam-splitter (BS) interaction $a_1^{\dag}b_1+a_1b_1^{\dag}$ on resonance (blue-detuning), when the detuning frequency $|\Delta|<2\omega_m$.

According to the above analysis, the bistable behavior can be easily induced under the condition of weak dissipation, strong driving, weak single-photon coupling strength and small detuning.
Same requirements are also needed in quantum information processing, such as ultrasensitive detection \cite{PhysRevA.94.013808},  radiation-pressure cooling \cite{nature.475.359} and optomechanical entanglement \cite{PhysRevA.94.053807}.
This applications also need weak dissipation to reduce the noise of the system, strong driving to enhance the linear coupling rate and small detuning to increase the strength of interaction.
Therefore, the system's unstable behavior must be considered in the study of quantum effects.

\section{Optomechanical cooling in instable regime}
\begin{figure}[tb]
  \centering
  \includegraphics[width=8.8cm]{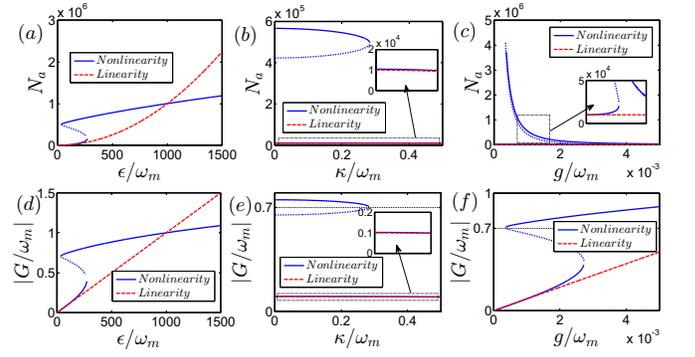}\\
  \caption{Typical curves for the mean-field cavity occupation $N_a$ and field enhanced coupling rate $G$ as a function of driving strength $\epsilon$ (a), (d); dispassion rate $\kappa$ (b), (e) and single-photon coupling rate $g$ (c), (f).
  In (a) and (d), the dispassion rate $\kappa/\omega_m=0.1$, single-photon coupling rate $g/\omega_m=10^{-3}$.
  In (b) and (e), the driving strength $\epsilon/\omega_m=100$, single-photon coupling rate $g/\omega_m=10^{-3}$.
  In (c) and (f), the dispassion rate $\kappa/\omega_m=0.1$, driving strength $\epsilon/\omega_m=10^2$.
  Other parameters are $\Delta=\omega_m$, $\gamma/\omega_m=10^{-4}$.}\label{fig3}
\end{figure}

In sideband cooling, the dynamics of the system can be linearized when the coupling rate is weak enough.
Under this condition, we can substitute the mean-field steady solutions $\alpha_0$, $\langle x_0\rangle$ into the quantum part and then get the final occupation number of the phonons after cooling.
According to the above analysis (see \fref{fig2}), the system can exhibit strong classical nonlinearity with the weak dispassion rate $\kappa$ and strong driving strength $\epsilon$, even that the coupling rate is still in the weak coupling regime $g\ll \omega_m$.
Under this condition, the same parameters may have different cooling behavior.
To study cooling effect in the nonlinear regime, we first consider the classical part of the system i.e. \esref{cpart}.
In the steady state they read,
\begin{eqnarray}
  0 &=& -(i\Delta+\kappa/2)\alpha+ig\alpha(\beta+\beta^*)-i\epsilon,\nonumber\\
  0 &=& -(i\omega_m+\gamma/2)\beta+ig|\alpha|^2.
\end{eqnarray}
Combining the above equations we obtain a third-order polynomial root equation for the mean-field cavity occupation,
\begin{eqnarray}
  0 &=& 4 \chi^2 n^3-4 \Delta \chi n^2+(\Delta^2+\kappa^2/4)n-\epsilon^2,\label{mfen}
\end{eqnarray}
where $n=|\alpha|^2$ is the mean photon number in the cavity, and $\chi=g^2/(\omega_m+\gamma^2/4\omega_m)$ is the nonlinear parameter.
This equation is commonly used to discuss the bistability of the optomechanical system \cite{PhysRevA.88.043826,PhysRevA.49.1337}.
Equation (\ref{mfen}) indicates that the mean-field equations have either one or three solutions depending on the number of real roots of the polynomial.
While the mechanical quality factor $Q_m=\omega_m/\gamma$ is much larger than 1, the nonlinear parameter $\chi$ approximately equals to $g^2/\omega_m$.
The three roots depend on the detuning $\Delta$, cavity dissipation rate $\kappa$, single-photon coupling rate $g$ and driving power $\epsilon$.
The effect of these four parameters on stability has been analyzed in \fref{fig2}.
A similar observation can be made concerning the mechanical resonator,
\begin{eqnarray}
  0 &=& \omega_m g^2 x^3-2g \Delta \omega_m x^2\nonumber\\
  &&+(\Delta^2+\kappa^2/4)\omega_m x+2g \epsilon^2,\label{mfex}
\end{eqnarray}
where $x=(\beta^*+\beta)$ is the mean displacement of the oscillator.
It is obvious that \eref{mfex} has also one or three roots.
This can be used to study the bistable behavior of oscillator equilibrium displacements \cite{PhysRevA.88.043826}.
For the discussion of the quantum part, the field-enhanced coupling rate $G=\alpha g$ is a more direct parameter.
Therefore, we focus mainly on the bistability of the average photon number $n=|\alpha|^2$ i.e. \eref{mfen}.
And the coupling rate $G$ is a function of $\kappa$, $g$, $\epsilon$, $\gamma$ under the consideration of nonlinear effect, i.e. $G(g,\kappa,\epsilon,\gamma)$.
Following consideration of the experimental situation, we plot the mean photon number $N_a$ and the field enhanced coupling rate $G$ as a function of parameters $\epsilon$, $\kappa$ and $g$ in \fref{fig3} in single photon weak-coupling regime.
In order to highlight the nonlinear effects of the system, we compare the condition which includes nonlinear term $ig\alpha(\beta+\beta^*)$ (CNL) with the one that does not (CL).

As shown in \fref{fig3}(a), under nonlinear condition (CNL), if the driving strength $\epsilon$ is weak enough, the curve is the same as the linear condition (CL).
As the driving strength increases until $\epsilon>35\omega_m$, the dynamic of the system will exhibit bistable behavior which have two different stable state with the same value of $\epsilon$ (one is metastable \cite{PhysRevA.84.033846}, blue dotted line).
Under this condition, one of the mean photon state $N_a$ is almost the same as the linear condition, but the other is much higher than the linear condition.
The bistable behavior of the system disappears (there is only one steady solution) following with the continuous increases of the driving strength.
When the driving strength is higher than a specific value $10^3\omega_m$, the mean photon number in linear condition will be higher than the one that in the nonlinear condition.
The same dynamic behavior of $G$ can be also seen in the \fref{fig3}(d).
As shown in \fref{fig3}(b), the mean photon number $N_a$ decreases when the dissipation $\kappa$ increases (red dashed line).
For nonlinearity condition, the system exhibits bistable behavior when the dissipation rate is weaker than a specific value $0.28\omega_m$. One of the mean photon state is almost the same as the linear condition, but the other is much higher than the linear condition.
Analogous dynamic behavior of $G$ can be also seen in \fref{fig3}(e), and $G$ will appear with a large value grater than $0.7 \omega_m$ in nonlinear condition.
In the nonlinearity regime, the system exhibits bistable behavior when the single-photon coupling rate achieves a specific threshold $3.5\times 10^{-4}\omega_m$.
As shown in \fref{fig3}(c) (blue solid line), one of the mean photon state $N_a$ is almost the same as the linear condition, but the other is much higher than the linear condition.
As the coupling strength $g$ increases, the bistable behavior of the system disappears (there is just one steady solution), and the dynamic behavior of the system is consistent with the linearity condition.
Analogous dynamic behavior of $G$ can be also seen in the \fref{fig3}(f).
Thus, in bistable regime, the system can achieve higher average photon number $N_a$ and larger coupling rate $G$ when $\epsilon>35\omega_m$, $\kappa<0.28\omega_m$, $g>3.5\times 10^{-4}\omega_m$ for given parameters.
Under the linear condition, the ultra-strong driving, high quality factor and strong coupling rate are needed to achieve the same effect on the average photon number and field enhanced coupling rate.
In optomechanical system, according to \esref{qpart}, the coupling rate $G(g,\kappa,\epsilon,\gamma)$ and the dissipation rate $\kappa$ are the key parameters of the evolution of quantum part.
Therefore, we can use adjustable parameters to modulate the dynamic characteristics of the system.

\subsection{Sideband cooling in quantum nonlinearity regime}
In order to facilitate the analysis and calculation, we split the optical and mechanical field operators into the classical and quantum components, and the corresponding dynamical equations of the system can be also divided into classical and quantum nonlinear equations, i.e. \esref{cpart} and \esref{qpart}.
In equations (\ref{qpart}), the nonlinear terms $iga_1(b_1+b_1^{\dag})$ and $ig a_1^{\dag}a_1$ directly depend on the key parameter single-photon coupling rate $g$.
When the single-photon coupling rate $g$ is rather small, nonlinear terms can be neglected as infinitesimal of high order.
But as $g$ increases, the nonlinearity of the system should be considered.
Therefore, it is necessary for us to study what is the specific coupling range that can ignore the nonlinear terms and how the influence of the quantum nonlinearity on the sideband cooling.
According to \esref{cpart} and \esref{qpart}, by using the master equation under the reasonable truncation of the operators we can obtain the solutions of steady state phonon number $N_{bs}$ with and without nonlinear terms $iga_1(b_1+b_1^{\dag})$ and $ig a_1^{\dag}a_1$ (details see Appendix A).
As shown in \fref{fig4}, with the increase of the single-photon coupling rate $g$, the number of the steady state phonon number $N_{bs}$ will increase rapidly in a very small intervals and then changes slowly around $3.3$.
In the region $g/\omega_m<3\times 10^{-3}$, as shown in the subgraph in \fref{fig4}, the steady state phonon number shows remarkable bistable behavior.
Under this condition, the nonlinear effect from quantum part can be neglected, and the evolution curve containing quantum nonlinearity (red-solid line QNL) agrees perfectly with the one not containing (blue-dashed line QL).
The nonlinear dynamical behavior of the steady-state phonon number depends mainly on the nonlinearity of the classical part i.e. \esref{cpart}.
With the increase of the single-photon coupling rate, the quantum nonlinear effect gradually emerges.
When the coupling rate $g$ is approximately over $0.015\omega_m$, as shown in the subgraph in \fref{fig4}, there are obvious differences between curves QL and QNL.
Under this condition, we can not ignore the influence of quantum nonlinearity on the sideband cooling.
Up to now, most experimental realizations of cavity optomechanics are still in the single-photon weak coupling limit \cite{RevModPhys.86.1391}, i.e., $g/\omega_m\ll1$.
The proper choice of the single-photon coupling rate $g$ is in order of magnitude $10^{-3}$ or even smaller.
In addition, our previous analysis indicates that bistability can also occur even with a small value of $g$.
In this case, the linearization description of the quantum part is valid, so the quantum nonlinearity can be neglected.
It is worth noting, when we introduce the nonlinear terms $iga_1(b_1+b_1^{\dag})$ and $ig a_1^{\dag}a_1$, the steady state of the oscillator maybe a complex non-Gaussian state.
Under this condition, the fluctuation of the phonon number can be no longer described by a specific temperature. That is to say, we can not use the thermal temperature to evaluate the cooling effect.
Nevertheless, based on the purpose of suppressing quantum fluctuations in mechanical cooling, the steady-state fluctuation of phonon number $N_{bs}$ can be still used as a standard to indicate the cooling effect.
\begin{figure}
  \centering
  \includegraphics[width=8.5cm]{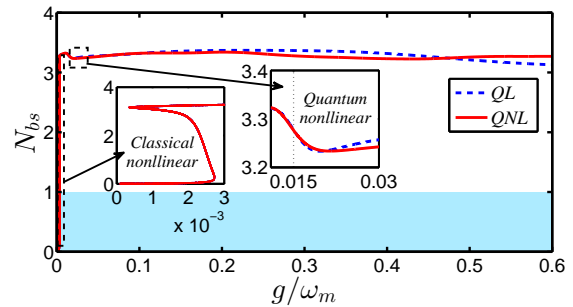}\\
  \caption{The comparison of the steady state cooling effect with quantum nonlinearity (QNL) and without quantum nonlinearity (QL).
  The dispassion rate $\kappa/\omega_m=0.1$, the thermal excitation number $n_{th}=200$, driving strength $\epsilon/\omega_m=100$, single-photon coupling rate $g/\omega_m=10^{-3}$, other parameters are the same with \fref{fig3}.}\label{fig4}
\end{figure}
\subsection{Sideband cooling in quantum linearity regime}
\begin{figure}
  \centering
  \includegraphics[width=5cm]{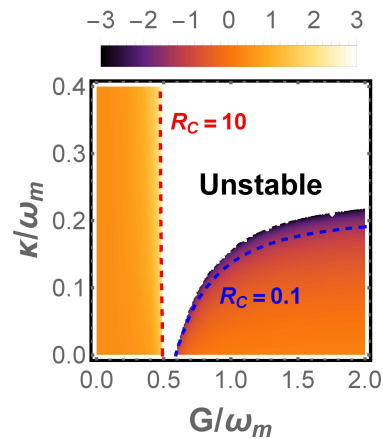}\\
  \caption{$log_{10}(\mathcal{R}_{C})$ as a function of $\kappa$ and $G$.
  The red-dashed line denotes $\mathcal{R}_{C}=10$, and the blue-dashed line denotes $\mathcal{R}_{C}=0.1$.
  The white area denotes the unstable area.
  Other parameters are same with \fref{fig4}.
  }\label{figadd}
\end{figure}
\begin{figure}[tb]
  \centering
  \includegraphics[width=8.5cm]{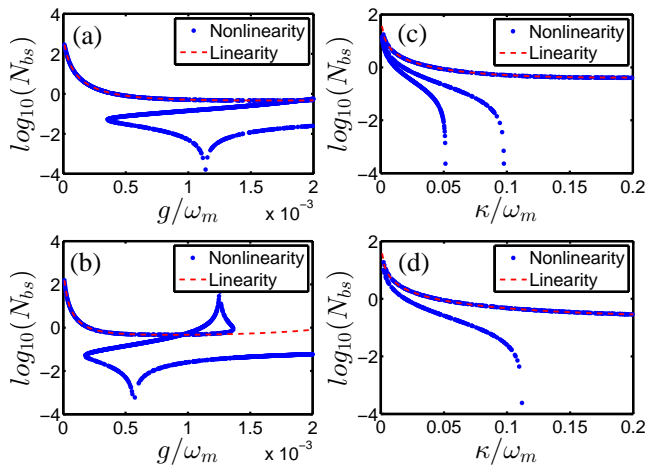}\\
  \caption{Comparison of cooling effect under linear and nonlinear conditions.
  (a) and (b) indicate the occupation phonon number $N_{bs}$ as a function of $g$.
  (c) and (d) indicate the occupation phonon number $N_{bs}$ as a function of $\kappa$.
  The cavity dissipation rate of (a) and (b) are $\kappa/\omega_m=0.1$.
  The single-photon coupling rate of (c) and (d) are $g/\omega_m=10^{-3}$.
  The driving strength of (a) and (c) are $\epsilon/\omega_m=100$.
  The driving strength of (b) and (d) are $\epsilon/\omega_m=200$.
  Other parameters are same with \fref{fig3}.
  }\label{fig5}
\end{figure}
After linearization (QL), we can derive the dynamical equations of mean values of the second-order moments according to \esref{qpart} (see Appendix B).
Under blue sideband condition $\Delta=\omega_m$, we can completely solve the differential equations for the mean values of the second-order moments $\langle a_1^{\dag} a_1\rangle$, $\langle b_1^{\dag} b_1\rangle$, $\langle a_1^{\dag} b_1\rangle$, $\langle a_1 b_1\rangle$, $\langle a_1^2\rangle$, $\langle b_1^2\rangle$ and obtain the steady state phonon number of sideband cooling $N_{bs}$ by setting $t \rightarrow \infty$.
Under the condition $4|G|^2\gg\kappa \gamma$, we can obtain an approximating expression for the steady state phonon number $N_{bs}$.
Similar approximating solution is also used in the discussion of dissipative cooling \cite{PhysRevLett.110.153606}.
\begin{eqnarray}
  N_{bs} &\approx& N_{RW}+N_{ARW},
\end{eqnarray}
where $N_{RW}=(4|G|^2+\kappa)n_{th}/[4|G|^2(\kappa+\gamma)]$ denotes the contribution from the rotating-wave effect i.e. $\langle a_1^{\dag} a_1\rangle$, $\langle b_1^{\dag} b_1\rangle$, $\langle a_1^{\dag} b_1\rangle$,
and $N_{ARW}=[4\omega_m^2(\kappa^2+8|G|^2)+\kappa^2(\kappa^2-8|G|^2)]/16\omega_m^2(4\omega_m^2+\kappa^2-16|G|^2)$ denotes the contribution from the anti-rotating-wave effect i.e. $\langle a_1 b_1\rangle$, $\langle a_1^2\rangle$, $\langle b_1^2\rangle$ (details see Appendix B).
In general, the anti-rotating-wave terms will induce the double excitation processes, which will heat the phonons and even cause instability of the system \cite{PhysRevLett.99.093901}.
Therefore, in many cooling protocols, it is desirable to avoid the anti-rotating-wave effect as far as possible \cite{PhysRevLett.99.093901,PhysRevLett.110.153606}.
We note that the anti-rotating-wave terms does not only cause negative effect on sideband cooling, but also can induce double quantum annihilation effects to optimize cooling, such as $b_1b_1$ and $a_1 b_1$.
To investigate sideband cooling better, we define cooling factor $\mathcal{R}_C=N_{bs}/N_{RW}$ in which $\mathcal{R}_C>1$ indicates that the anti-rotating wave effect heats the mechanical oscillator and $0\leq\mathcal{R}_C<1$ indicates that the anti-rotating-wave effect cools the mechanical oscillator.
As shown in \fref{figadd}, when $G/\omega_m\leq0.5$, the cooling factor $\mathcal{R}_{C}$ is greater than 1 and increases with the increase of coupling rate $G$.
When the parameters exceeds the red-dashed line region ($\mathcal{R}_{C}=10$), the cooling factor increases rapidly and tends to be unstable very quickly.
In this case, anti-rotating wave terms exhibit negative effect for sideband cooling.
While $G/\omega_m\geq0.6$, the cooling factor $\mathcal{R}_{C}\leq1$, especially from the blue-dashed line ($\mathcal{R}_{C}=0.1$) to the dark colored zone, the cooling factor drops sharply and is far less than 1.
Therefore, if the coupling rate $G$ is large enough, the anti-rotating-wave terms will be beneficial to the sideband cooling under the premise that the system has steady-state solutions.

We plot the steady state cooling effect as a function of single-photon coupling rate $g$ and cavity dissipation rate $\kappa$ with different driving strength $\epsilon$ in \fref{fig5}.
Under nonlinear condition (CNL), the steady state cooling phonon number has obvious valley value, which is much lower than that in the linear case (CL).
As shown in \fref{fig5}(a), when the single-photon coupling rate $g$ is relatively small, the steady state phonon number $N_b$ has only one stable solution, and the nonlinear curve agrees with the linear case.
With the increase of coupling rate, steady state phonon number will appear bistability, which is consistent with the results that has been discussed in \fref{fig3}(f).
One of the stable solution is still consistent with the linear case, while the other one is much lower, and it will reach a minimum value $N_b\approx 10^{-4}$ at $g/\omega_m=1.13\times 10^{-3}$.
In combination with \fref{fig3}(f), we can find that at the minimum point of the phonon number, the field-enhanced coupling rate $G$ approximately equals to $0.8\omega_m$.
Then according to the dissipation rate we selected $\kappa/\omega_m=0.1$, we find that the corresponding cooling factor $\mathcal{R}_C$ is exactly in the lowest region in \fref{figadd}.
That is to say, the effect of the anti-rotating-wave terms will optimize the cooling effect in this case.
In \fref{fig5}(b), we let the driving strength $\epsilon$ equal to two times of that in \fref{fig5}(a).
Through the comparison between the two figures, we find that the stronger the driving strength is, the smaller the single-photon coupling is required to achieve bistability effect and optimal cooling.
In addition, in the high value region of single-photon coupling rate, there is only one stable solution existing in the nonlinear condition, and the cooling effect is also lower than the one that in the linear case.

As shown in \fref{fig5}(c) and (d), in the low dissipation region ($\kappa/\omega_m< 0.1$ and $\kappa/\omega_m< 0.12$), there is obvious bistability of the steady state phonon number.
In \fref{fig5}(c), one of the coexisting steady states is consistent with the linear condition, and the other have a minimum value at the dissipation rate $0.1\omega_m$.
Similarly, according to \fref{figadd}, the optimal cooling is resulting from the bistability which due to the additional cooling effect of the anti-rotating-wave terms.
When the dissipation rate is greater than $0.1\omega_m$, the bistable effect disappears, and the curve agrees with the linear case, and the commonly used classical linear approximation is valid.
Through the comparison between \fref{fig5}(c) and \fref{fig5}(d), we find that stronger driving strength allows the system to achieve optimal cooling effects at larger cavity dissipation rate.
Therefore, we can achieve a much lower steady state cooling phonon number by using the nonlinearity of the system in optomechanical cooling.
In order to utilize the nonlinearity of the system to optimize cooling, under the given parameters, we need to select the appropriate single-photon coupling rate (it is not as strong as possible) and keep the dissipation rate $\kappa/\omega_m$ less than specific value, $\kappa/\omega_m<0.1$ for \fref{fig5}(c), which can be easily achieved in existing experiments.

\section{Discussion and conclusion}
The realization of single-photon coupling rate in optomechanical systems are still in weak coupling regime under current experimental conditions i.e., $g/\omega_m\ll1$.
Therefore, for the sideband cooling of the optomechanical system, the linear coupling term $(G^*a_1+G a_1^{\dag})(b_1+b_1^{\dag})$ still plays an important role.
While the quantum nonlinear effects such as muti-photon and muti-phonon interaction are relatively weak \cite{PhysRevA.85.051803,PhysRevA.92.023811}, which have been analyzed in section III-A.
The fundamental reason for the sideband cooling optimization of bistability behavior is the anti-rotating wave effect caused by the classical nonlinearity.
In our result, the system can exhibit strong nonlinear effects in the unstable region even the nonlinear interaction of system is weak.
Considered with present experimental conditions, using bistable effect is a meaningful choice to optimize sideband cooling.
In our analysis, to obtain optomechanical bistability, we only need $g/\omega_m>3.5\times 10^{-4}$ and $\kappa/\omega_m<0.28$ at the appropriate driving strength, which can be easily realized in the nanobeam optomechanical resonator cavity \cite{Nature10461}.
Where the mechanical frequency $\omega_m/2\pi=3.68GHz$, optical linewidth $\kappa/2\pi=500MHz$ and single-photon coupling rate $g/2\pi=910KHz$.
In addition the switching between the two bistable states can be realized by the injected signal \cite{ncomms15141}.

In conclusion, we present a scheme to optimize the sideband cooling by using the bistability of the optomechanical system.
By using RH criterion to study the stability of the system, we find that the system can exhibit unstable behaviors caused by the classical nonlinearity, which do not require strong drive, strong single-photon coupling and high cavity Q factor.
Then by investigating the bistable behavior of the system, we find that there is an obvious parameter amplification of the field-enhanced coupling rate $G$.
We also study the steady state cooling of the mechanical oscillator with and without quantum nonlinearity respectively.
It is obvious that cooling optimization mainly depends on the classical nonlinear in weak single-photon coupling regime.
Compared with the original linearized cooling scheme \cite{RevModPhys.86.1391} in the stable regime, our scheme can achieve a lower steady state cooling phonon number $N_{bs}$ and has no requirement for strong single-photon coupling $g$ and high cavity quality factor $Q$.
This provides a theoretical platform for optimizing oscillator cooling and studying nonlinear behavior of optomechanical systems.

\section*{ACKNOWLEDGMENTS}
We would like to thank Dr. Lei-Ming Zhou and Hui-Hui Qin for helpful discussions.
This work was supported by the Fundamental Research Funds for the Central Universities under Grant No. 2015MS55.\\

\appendix
\section{Dynamic of mechanical oscillator with quantum nonlinearity}
The optical and mechanical field operators can be split into the classical and quantum components: $a\rightarrow\alpha+a_1$ and $b\rightarrow\beta+b_1$, where $\alpha\equiv\langle a\rangle$, $\beta\equiv \langle b \rangle$. In the rotating frame, the corresponding Hamiltonian \eref{hmi} can be rewritten as $H=H_c+H_q$. Where $H_c$ denotes the mean part of the Hamiltonian, which dominates the dynamical evolution of the classical components \esref{cpart}.
$H_q$ denotes the fluctuation part of the Hamiltonian, which dominates the dynamical evolution of the quantum components \esref{qpart}. We have
\begin{eqnarray}
  H_q &=& \Delta'a_1^{\dag}a_1+\omega_mb_1^{\dag}b_1\\
  &&-(G^*a_1+G a_1^{\dag})(b_1+b_1^{\dag})-g a_1^{\dag}a_1(b_1+b_1^{\dag}),\nonumber \label{hq}
\end{eqnarray}
where $\Delta'=\Delta-g(\beta+\beta^*)$ is the detuning modified by the optomechanical coupling and $G=\alpha g$ describes the linear coupling strength.
We see that the time-dependent coefficients $\Delta'$ and $G$ are determined by $\alpha$ and $\beta$.
That is, the classical nonlinear dynamics is manifested in the quantum properties of the system.
We derive the dynamic solutions of the classical parts by solving \esref{cpart}.
Substituting the solution of $\alpha$ and $\beta$ into \eref{hq}, we can obtain the parameter modulated Hamiltonian of the fluctuation part.
Then, according to the parameter modulated Hamiltonian, the average phonon fluctuation can be formally given by $N_b=Tr[b_1^{\dag}b_1\rho]$. Where $\rho$ is the density operator of system. The evolution of the density operator $\rho$ for the Hamiltonian $H_q$ can be described by the master equation,
\begin{eqnarray}
  \dot{\rho} &=& -i[H_q,\rho]\\
  &&+\frac{\kappa}{2}\mathcal{D}[a]\rho+\frac{\gamma}{2}(n_{th}+1)\mathcal{D}[b]\rho+\frac{\gamma}{2}n_{th}\mathcal{D}[b^{\dag}]\rho, \nonumber
\end{eqnarray}
where $n_{th}=[exp(\omega_m/k_B T_M-1)]^{-1}$ is the average thermal occupancy number of the oscillator, and $\mathcal{D}[o] = 2 o \rho o^{\dag}-o^{\dag}o \rho - \rho o^{\dag}o$ is the Lindblad dissipation superoperator.
When the excitation of the fluctuation part is small, we can solve the system by truncating the dimension of the operators.
Thus we can obtain the steady-state cooling phonon fluctuations number $N_{bs}$, which are governed by a parameters $\alpha$ and $\beta$.

\section{Dynamic of mechanical oscillator without quantum nonlinearity}
Under strong driving and weak coupling condition, the classical components dominate and the nonlinear terms $g a_1^{\dag}a_1(b_1+b_1^{\dag})$ can be neglected.
Since the Hamiltonian is linear, it does not mix moments with different orders.
To calculate the mean phonon number, we need to determine the mean values of all the second-order moments, the differential equations are given as
\begin{widetext}
\begin{eqnarray}\label{eqadd}
  \frac{dN_a}{dt} &=& -\kappa N_a-i G^*(\langle a_1 b_1\rangle+\langle a_1^{\dag} b_1\rangle^*)+iG(\langle a_1^{\dag} b_1\rangle+\langle a_1 b_1\rangle^*),\nonumber \\
  \frac{dN_b}{dt} &=& -\gamma N_b+\gamma n_{th}-i(G\langle a_1^{\dag} b_1\rangle+G^*\langle a_1 b_1\rangle-G\langle a_1 b_1\rangle^*-G^*\langle a_1^{\dag} b_1\rangle^*),\nonumber\\
  \frac{d\langle a_1^{\dag} b_1\rangle}{dt} &=&-[ i(\omega_m-\Delta)+(\kappa+\gamma)/2]\langle a_1^{\dag} b_1\rangle-i(G^*\langle b_1^2\rangle+G^*N_b-G\langle a_1^2\rangle^*-G^*N_a),\nonumber\\
  \frac{d\langle a_1 b_1\rangle}{dt} &=&-[ i(\omega_m+\Delta)+(\kappa+\gamma)/2]\langle a_1 b_1\rangle+i(G\langle b_1^2\rangle+GN_b+GN_a+G+G^*\langle a_1^2\rangle),\nonumber\\
  \frac{d\langle a_1^2\rangle}{dt} &=&-(2i\Delta+\kappa)\langle a_1^2\rangle+2iG(\langle a_1 b_1\rangle+\langle a_1^{\dag} b_1\rangle^*),\nonumber\\
  \frac{d\langle b_1^2\rangle}{dt} &=&-(2i\omega_m+\gamma)\langle b_1^2\rangle+2i(G^*\langle a_1 b_1\rangle+G\langle a_1^{\dag} b_1\rangle).
\end{eqnarray}
Different from the usual linearization method \cite{PhysRevLett.110.153606}, $G$ is dependent on the parameters $g$, $\epsilon$, $\kappa$ and $\gamma$.
Where the steady-state covariance matrix is used to obtain the final occupancy of the mechanical resonator $N_{bs}$.
We note that in the above calculation, cut-off of the density matrix is not necessary and the solutions are exact.

In the stable regime, the system finally reaches the steady state, and the derivatives in \esref{eqadd} all become zero.
Then the second-order moments in the steady state satisfy a set of algebraic equations.
Under the condition $\Delta'=\omega_m$ and cooperativity $4|G|^2\gg\kappa \gamma$, we obtain the final phonon occupancy
\begin{eqnarray}
  N_{bs} &=& \frac{(4|G|^2+\kappa)n_{th}}{4|G|^2(\kappa+\gamma)}+\frac{4\omega_m^2(\kappa^2+8|G|^2)
  +\kappa^2(\kappa^2-8|G|^2)}{16\omega_m^2(4\omega_m^2+\kappa^2-16|G|^2)}.\label{nbarw}
\end{eqnarray}
In addition, under the rotating-wave approximation,we can set $\langle a_1 b_1\rangle$, $\langle a_1^2\rangle$ and $\langle b_1^2\rangle$ zero and then we can obtain the final phonon occupancy
\begin{eqnarray}
N_{bs}&=&\frac{(4|G|^2+\kappa)n_{th}}{4|G|^2(\kappa+\gamma)}.\label{nbrw}
\end{eqnarray}
By comparing \eref{nbarw} with (\ref{nbrw}), we can find that the first term of \eref{nbarw} indicates the
contribution from the rotating-wave effect, the second term of \eref{nbarw} indicates the contribution from the anti-rotating-wave effect.

\end{widetext}
\bibliography{mscl}
\end{document}